\documentclass[aps,
floatfix,superscriptaddress,a4paper]{revtex4-1}
\usepackage{graphicx,amsmath,bbm,mathrsfs,amssymb,psfrag}

\evensidemargin \oddsidemargin
\begin{document}

\author{Laure Gouba\\
The Abdus Salam International Centre for
Theoretical Physics (ICTP),\\
 Strada Costiera 11,
I-34151 Trieste Italy. \\
Email: lgouba@ictp.it}

\title{Time-dependent q-deformed bi-coherent states for generalized uncertainty relations}

\begin{abstract}
We consider the time-dependent bi-coherent states that are essentially 
the Gazeau-Klauder coherent states for the two dimensional noncommutative harmonic oscillator.
Starting from some q-deformations of the oscillator algebra for which the entire 
deformed Fock space can be constructed explicitly, we define
the q-deformed bi-coherent states. We verify  the generalized Heisenberg's uncertainty relations 
projected onto these states. For the initial value in time the states are shown 
to satisfy a generalized version of Heisenberg's uncertainty relations. For the initial value in 
time and for the parameter of noncommutativity $\theta =0$, the inequalities are saturated 
for the simultaneous measurement of the position-momentum observables.
When the time evolves the uncertainty products are different from their values at the initial time 
and do not always respect the generalized uncertainty relations.
\end{abstract}

\maketitle

\tableofcontents

\section{Introduction}
Quantum harmonic oscillator is one of the potential tools that one needs 
to handle the quantum world. In standard quantum mechanics, it can be described 
in terms of the creation and annihilation operators which obey a commutation relation 
$a a^\dagger -a^\dagger a =1$.  The q- deformations of this algebra were introduced 
independently by Biedenharn \cite{bieden} and Macfarlane \cite{alan}. Later several 
definitions and investigations of q-oscillators appeared in the literature, \cite{fu},\cite{dama},
\cite{delga1}, \cite{delga2}, \cite{max}, 
\cite{odz},\cite{quesn},\cite{hounk}. 
The algebras satisfied by the canonical variables resulting from q-deformed oscillator 
algebras have been shown to be related to noncommutative spacetime structures leading 
to minimal lengths and minimal momenta as a result of a generalized 
version of Heisenberg'uncertainty relations \cite{bijan},\cite{andreas},\cite{laure}, 
\cite{dey1}. The existence of explicit states satisfying these relations and their 
construction have been investigated in \cite{dey}. Using the Gazeau-Klauder coherent 
states \cite{gazeau}, they have shown for a noncommutative harmonic oscillator to first 
order perturbation theory in the deformation parameter that these states not only satisfy 
the generalized uncertainty relations, but even saturate them at all times.
An extension of this analysis is the study of time-dependent q-deformed coherent 
states for generalized uncertainty relations \cite{fring} in which the properties of generalized 
time-dependent q-deformed coherent states for a noncommutative harmonic oscillator have 
been investigated. The states are shown to satisfy a generalized version of the 
Heisenberg's uncertainty relations and the inequalities are saturated for the initial 
value in time. 

The aim of this paper is to extend 
the analysis in \cite{fring} to higher dimension. In order to do 
that we revisit the formulation and interpretation 
of noncommutative quantum mechanics \cite{scholtz}, where the formalism has been 
applied successfully to the case of two dimensional harmonic oscillator 
that is described by creation and annihilation operators $a_i, a_i^\dagger, \; i=1,2 $ obeying
the commutation relation $a_ia_j^\dagger - a_ja_i^\dagger = \delta_{ij},\; i, j = 1,2$. Those operators
have been useful in constructing displacement operators and gaussian states in the study 
the classical limits of quantum mechanics on a non-commutative configuration 
space \cite{fabio}. In this manuscript we q-deform the algebra that they obey 
and we obtain a two dimensional 
q-deformed oscillator algebra and then we define the corresponding q-deformed version of the Fock space from 
which the q-deformed bi-coherent states are constructed. By bi-coherent states, we refer 
to the ones defined by S. T. Ali and F. Bagarello \cite{ali} which can be considered as multidimensional 
Gazeau-Klauder coherent states \cite{jpnov}.

Our manuscript is organized as follow: in section (\ref{sec1})
we revisit the two dimensional harmonic oscillator studied in \cite{scholtz}. In section 
(\ref{sec2}) we define the q-deformed Bi-coherent states. The generalized uncertainties are discussed 
in (\ref{sec3}) and concluding remarks are given in (\ref{sec4}).

\section{The Harmonic oscillator in noncommutative configuration space revisited }\label{sec1}
We shortly review the formalism of noncommutative quantum mechanics, more details being 
available in \cite{scholtz}. We consider two dimensional noncommutative configuration space,
where the coordinates satisfy the commutation relation
\begin{equation}\label{eq1}
 \left[ \hat x_i, \hat x_j\right] = i\theta\epsilon_{ij},
\end{equation}
 with $\theta$ a real positive parameter and $\epsilon_{i,j}$ the
completely antisymmetric tensor with $\epsilon_{1,2} = 1$.
Since, the operators
\begin{equation}
\label{Bop}
b = \frac{1}{\sqrt{2\theta}}(\hat x_1 + i\hat x_2)\ ,\quad
b^\dagger = \frac{1}{\sqrt{2\theta}}(\hat x_1 -i\hat x_2)
\end{equation}
satisfy the commutation relations $[b, b^\dagger] =  1$, one can introduce
a Fock-like vacuum vector $\vert 0\rangle$ such that
$b\vert 0\rangle=0$ and construct a non-commutative
configuration space isomorphic to the boson Fock space
\begin{equation}
 \mathcal{H}_c = \textrm{span}\{|n\rangle \equiv \frac{1}{\sqrt{n!}}
(b^\dagger)^n |0\rangle\}_{n=0}^{n = \infty},
\end{equation}
where the span is taken over the field of complex numbers.

A proper Hilbert space over such non-commutative configuration space
is the Hilbert-Schmidt Banach algebra $\mathcal{H}_q$ of bounded operators
$\psi(\hat x_1, \hat x_2)\in \mathcal{B}(\mathcal{H}_c)$ on $\mathcal{H}_c$ such that
\begin{equation}
\label{eq4}
tr_c(\psi(\hat x_1, \hat x_2)^\dagger\psi(\hat x_1, \hat x_2))< \infty \ .
\end{equation}
The $\textrm{tr}_c$ denotes the trace over non-commutative configuration space and
$\mathcal{B}(\mathcal{H}_c)$ the set of bounded operators on $\mathcal{H}_c$.
This space has a natural inner product and norm
\begin{equation}\label{eq5}
 (\phi(\hat x_1, \hat x_2),\psi(\hat x_1, \hat x_2)) =
\textrm{tr}_c(\phi(\hat x_1, \hat x_2)^\dagger\psi(\hat x_1, \hat x_2)).
\end{equation}
Next we introduce the non-commutative Heisenberg algebra
\begin{eqnarray}
\label{NCHA}
 \left[\hat X_i,\:\hat P_j\right] = i\hbar\delta_{i,j},\quad
 \left[\hat X_i,\:\hat X_j\right] = i\theta\epsilon_{i,j},\quad
 \left[\hat P_i, \hat P_j\right] = 0,
 \end{eqnarray}
 where a unitary representation  in terms of the operators
$\hat X_i$ and $\hat P_i$ acting on the quantum Hilbert space $\mathcal{H}_q$ with
the inner product (\ref{eq5}) is
\begin{eqnarray}
 \hat X_i\psi(\hat x_1, \hat x_2) = \hat x_i\psi(\hat x_1,\hat x_2), \quad
 \hat P_i\psi(\hat x_1,\hat x_2) = \frac{\hbar}{\theta}\epsilon_{i,j}
 \left[\hat x_i,\: \psi(\hat x_1,\hat x_2)\right].
\end{eqnarray}
In the above representation, the position acts by left multiplication and
the momentum adjointly.
Let's consider the Hamiltonian of the two-dimensional non-commutative
harmonic oscillator
\begin{eqnarray}\label{e7}
 \hat H = \sum_{i =1}^2\left(\frac{1}{2m}\hat P_i^2 +
\frac{1}{2}m\omega^2\hat X_i^2 \right),
\end{eqnarray}
where the operators $\hat X_i, \hat P_i,\: i =1,2$ are acting on the
quantum Hilbert space (\ref{eq4}) and satisfy the commutation relations 
(\ref{NCHA}).

One can associate to the position and momentum operators, 
the operators 
$A_i,\; A_i^\dagger\;  i=1,2$
\begin{eqnarray}\label{ca}
 A_1 &=& \frac{1}{\sqrt{K_1}}
\left( -\frac{\lambda_1}{\hbar}\hat X_1 - i\hat P_1 -
i\frac{\lambda_1}{\hbar}\hat X_2 +\hat P_2 \right);\\
  A_1^\dagger &=& \frac{1}{\sqrt{K_1}}
\left( -\frac{\lambda_1}{\hbar}\hat X_1 + i\hat P_1
+i\frac{\lambda_1}{\hbar}\hat X_2 +\hat P_2 \right);\\
 A_2 &=& \frac{1}{\sqrt {K_2}}
\left( \frac{\lambda_2}{\hbar}\hat X_1 + i\hat P_1
-i\frac{\lambda_2}{\hbar}\hat X_2 +\hat P_2 \right);\\
 A_2^\dagger &=& \frac{1}{\sqrt{K_2}}
\left( \frac{\lambda_2}{\hbar}\hat X_1 - i\hat P_1 +
i\frac{\lambda_2}{\hbar}\hat X_2 +\hat P_2 \right),
\end{eqnarray}
where
\begin{eqnarray}\label{par1}
 \lambda_1 &=& \frac{1}{2}
\left( m\omega\sqrt{4\hbar^2 + m^2\omega^2\theta^2} + m^2\omega^2\theta\right);\;
K_1 = \lambda_1\left(4 + \frac{2\lambda_1\theta}{\hbar^2} \right);\\\label{par2}
\lambda_2  &=& \frac{1}{2}
\left( m\omega\sqrt{4\hbar^2 + m^2\omega^2\theta^2} - m^2\omega^2\theta\right);\;
K_2 = \lambda_2\left(4 - \frac{2\lambda_2\theta}{\hbar^2} \right).
\end{eqnarray}
They satisfy the algebra
\begin{equation}\label{aacc}
\left[A_i,\: \hat A_j^\dagger \right] 
= \delta_{ij};\:
\left[ A_i, \: \hat A_j \right] = 0,\; i,j = 0,1,
\end{equation}
and the equation
\begin{eqnarray}
 A_i\psi_{00} = 0,\quad i = 1,2;\;  \psi_{00} \equiv \vert 0,0\rangle.
 \end{eqnarray}
 admit a normalized solution 
\begin{equation}
\psi_{00}(\hat x_1,\hat x_2) = {\mathcal{N}}^{(-\frac{1}{2})}\exp(\frac{\alpha}{2\theta}(\hat x_1,\hat x_2)), 
\end{equation}
with
${\mathcal{N}}^{-1} = \frac{2\lambda_-}{\hbar^2}
\left(1-\frac{\theta\lambda_-}{2\hbar^2}\right)$ and 
$\alpha = \ln( 1-\frac{\theta}{\hbar^2}\lambda_-)
= -\ln(1 +\frac{\theta}{\hbar^2}\lambda_+)$.

The operators $A_i, A_i^\dagger, i= 1,2$ can then be interpreted 
as creation- annihilation operators. The model can be described by 
\begin{equation}
 \hat H = \frac{\lambda_1}{2m}(2 A_1^\dagger A_1 + 1) 
 +\frac{\lambda_2}{2m}(2 A_2^\dagger A_2 +1),
\end{equation}
 where the creation and annihilation operators $A_i, A_i^\dagger, i= 1,2$ 
 span the Fock space as follows
\begin{eqnarray}
A_1\vert n_1,n_2 \rangle &=&  n_1\vert n_1-1, n_2\rangle;\quad
A_1^\dagger \vert n_1,n_2 \rangle =  \sqrt{n_1+1}\vert n_1+1, n_2\rangle;\\
A_2\vert n_1,n_2 \rangle &=&  n_2\vert n_1, n_2-1\rangle; \quad 
A_2^\dagger \vert n_1,n_2 \rangle =  \sqrt{n_2+ 1}\vert n_1, n_2+ 1\rangle,
\end{eqnarray}
such that 
\begin{equation}
 A_1^\dagger A_1 \vert n_1,n_2\rangle = n_1\vert n_1,n_2\rangle; \quad 
 A_2^\dagger A_2 \vert n_1,n_2 \rangle = n_2 \vert n_1,n_2\rangle; \quad 
 \vert n_1,n_2 \rangle = \frac{(A_1^\dagger)^{n_1} (A_2^\dagger)^{n_2}}{\sqrt{n_1!n_2 !}}\psi_{00},
\end{equation}
and 
\begin{equation}
 H\vert n_1,n_2\rangle = \left(\frac{\lambda_1}{2m}(2n_1 +1) + \frac{\lambda_2}{2m}(2n_2 +1)\right)\vert n_1,n_2\rangle.
\end{equation}

\section{ q-deformed Bi-coherent states}\label{sec2}

Let us first shortly recall the Gazeau-Klauder scheme \cite{gazeau}. It is 
a method for constructing a real two-parameter set of coherent states 
$\{\vert J, \gamma\rangle \},\; J \ge 0$, and $-\infty < \gamma < +\infty$, 
associated to physical Hamiltonians $H$ which have discrete non-degenerate spectra. 
The states have to satisfy the following properties:
\begin{enumerate}
 \item Continuity: the mapping $(J,\gamma) \to \vert J, \gamma\rangle$ is continuous 
 in some appropriate topology.
 \item Resolution of unity : $\mathbb{I}= \int \vert J,\gamma\rangle \langle J,\gamma\vert d\mu(J,\gamma)$.
 \item Temporal stability: $e^{-iH t}\vert J,\gamma\rangle = \vert J,\gamma+\omega t\rangle$, $\omega = $ constant.
 \item Action identity: $\langle J, \gamma\vert H \vert J, \gamma\rangle = \omega J$.
\end{enumerate}
The construction of those states works if $H$ has no degenerate eigenstates and, 
furthermore, if the lowest eigenvalue is exactly zero. 

Let's consider a Hamiltonian $H$ with a discrete spectrum which is bounded below and 
adjusted so that $H \ge 0$, we assume in addition 
that the eigenstates are non-degenerate. The eigenstates $\vert n \rangle$ are orthonormal 
vectors satisfying 
\begin{equation}
 H\vert n \rangle = E_n\vert n\rangle, \quad n\ge 0,\quad 0= E_0 < E_1 < E_2 < \cdots , 
\end{equation}
where we set the eigenvalues $E_n = \omega \epsilon_n (=\hbar\omega\epsilon_n), \omega >0$ and fixed, with 
$0 =\epsilon_0 <\epsilon_1 < \epsilon_2 < \cdots$ being a sequence of dimensionless real 
numbers. 

The Gazeau-Klauder coherent states are 
defined as follows
\begin{equation}\label{gkcoh}
 \vert J,\gamma\rangle := \mathcal{N}(J)^{-\frac{1}{2}}\sum_{k =0}^\infty 
 \frac{J^{n/2} e^{-i\gamma \epsilon_n}}{\sqrt{\rho_n}}\vert n\rangle,
\end{equation}
where $J \ge 0$ and $\gamma\in \mathbb{R}$. The numbers $\rho_n$ are positive 
and are fixed by the requirement of the 
action identity to be $\rho_n = \epsilon_1\epsilon_2\ldots \epsilon_n$.
The normalization factor $\mathcal{N}(J)$, which turns out to be only
dependent on $J$, is chosen so that 
\begin{equation}
 \langle J,\gamma\vert J,\gamma\rangle = \mathcal{N}(J)^{-1}\sum_{n =0}^{+\infty}\frac{J^n}{\rho_n} \equiv 1, \quad 
 \textrm{then }\quad \mathcal{N}(J) \equiv \sum_{n =0}^{+\infty}\frac{J^n}{\rho_n}\,.
\end{equation}
  The 
domain of allowed $J,\, 0 \le J < R $, is determined by the radius of convergence 
$R$ in the series defining $\mathcal{N}(J)$. In the case of the harmonic oscillator, 
the equation (\ref{gkcoh}) is 
\begin{equation}
 \vert J,\gamma\rangle := e^{-\frac{J}{2}}\sum_{k =0}^\infty 
 \frac{J^{n/2} e^{-i n\gamma }}{\sqrt{n !}}\vert n\rangle,
\end{equation}
and setting $z = \sqrt{J}e^{-i\gamma}$ one finds the usual canonical coherent states 
\begin{equation}
 \vert z\rangle = e^{-\frac{1}{2}|z|^2}\sum_{n =0}^\infty\frac{z^n}{\sqrt{n!}}\vert n\rangle,
\end{equation}
also called the Glauber-Klauder-Sudarshan (GKS) states.

The multidimensional generalized coherent states have been presented for systems with several degrees of freedom 
\cite{jpnov} and also in \cite{ali} where the bi-coherent states have been defined.
The later approach inspires us in construction bi-coherent states for two dimensional 
noncommutative harmonic oscillator.
Since for the construction of the Gazeau-Klauder 
coherent states the eigenstates of the Hamiltonian should be non-degenerate and the 
lowest eigenvalue exactly zero, we consider the following operator  
\begin{equation}
 \tilde{H} = H - ((\lambda_1 + \lambda_2)/ 2m) \equiv \frac{\lambda_1}{m} A_1^\dagger A_1 + \frac{\lambda_2}{m} A_2^\dagger A_2.
\end{equation}
The Hamiltonian $\tilde{H}$ and $H$ have exactly the same dynamical content since they 
obey the same commutations relations with all the observables of the system.  In addition 
\begin{equation}
 \tilde{H}\vert n_1,n_2 \rangle =  (\frac{\lambda_1}{m}n_1+\frac{\lambda_2}{m}n_2)\vert n_1,n_2\rangle
\end{equation}
is such that the lowest eigenvalues of $\tilde{H}$ is zero while having 
common eigenvectors $\vert n_1,n_2\rangle$ 
with the Hamiltonian $H$. Next, we construct the bi-coherent states 
for the Hamiltonian $\tilde{H}$ as follows
\begin{equation}
 \vert J_1,\gamma_1; J_2,\gamma_2\rangle  = \frac{1}{\sqrt{\mathcal{N}(J_1,J_2)}} 
 \sum_{n_1,n_2}^{\infty}\frac{J_1^{\frac{n_1}{2}} J_2^{\frac{n_2}{2}}
 e^{i(\gamma_1n_1 +\gamma_2 n_2)}}{\sqrt{n_1 ! n_2 !}}\vert n_1,n_2\rangle,
\end{equation}
where $J_i\in \mathbb{R}_+, \gamma_i \in \mathbb{R},\; i= 1,2 $ and 
$\mathcal{N}(J_1,J_2)$ is the normalization factor.
As in \cite{ali}, the bi-coherent states have been shown in to be normalized to unity, 
to satisfy; the resolution of the identity, the temporal 
stability condition and the action identity. In the present case of Hamiltonian, the later stand as 
\begin{equation}
 e^{-i\tilde{H}t}\vert J_1,\gamma_1; J_2,\gamma_2\rangle =  
 \vert J_1, \gamma_1 +\frac{\lambda_1}{m}t;\; J_2, \gamma_2 + \frac{\lambda_2}{m}t\rangle,\quad
 \langle J_1,\gamma_1; J_2,\gamma_2\vert \tilde{H}\vert J_1,\gamma_1; J_2,\gamma_2\rangle = 
 \frac{\lambda_1}{m} J_1 + \frac{\lambda_2}{m}J_2.
\end{equation}
We q-deform the algebra (\ref{aacc}) by defining a new set of creation and annihilation operators 
$\tilde A_i,\; \tilde A_i^\dagger,\, i = 1,2$ satisfying 
\begin{equation}\label{daacc}
 \tilde A_i\tilde A_j^\dagger - \left((q^2-1)\delta_{ij}+1\right) \tilde A_j^\dagger \tilde A_i = \delta_{ij}, 
 \quad 0 < q\le 1;\quad [\tilde A_i,\;\tilde A_j] = 0 \; (i\neq j),
\end{equation}
The expression (\ref{daacc}) is the two-dimensional q-deformed oscillator algebra for the creation 
and annihilation operators $A_i,\;A_i^\dagger, i=1,2$, this deformation has minor different in 
the conventions with the ones defined in \cite{bieden},\cite{alan}, \cite{fu},\cite{dama}.
The corresponding Hamiltonian 
\begin{equation}
 \tilde{H}_q = \frac{\lambda_1}{m} \tilde A_1^\dagger \tilde A_1 + \frac{\lambda_2}{m} \tilde A_2^\dagger \tilde A_2.
\end{equation}
Next, we define the $q$-deformed analog of the Fock space involving $q$-deformed integer 
$[n]_q$ as 
\begin{equation}
 \vert n_1,n_2\rangle_q := \frac{(\tilde A_1^\dagger)^{n_1}(\tilde A_2^\dagger)^{n_2}}{\sqrt{[n_1]_q ![n_2]_q !}}\vert 0,0\rangle,
 \end{equation}
 where
 \begin{equation}
 [n_i]_q := \frac{1-q^{2n_i}}{1-q^2},\quad 
 [n_i]_q! := \Pi_{k=1}^{n_i}[k]_q,\quad
 \tilde A_i\vert 0,0\rangle = 0, \quad
 \langle 0,0\vert 0,0\rangle =1,\quad i=1,2.
\end{equation}
It follows that the operators $\tilde A_i^\dagger$ and $\tilde A_i$, $i =1,2$ act indeed 
as raising and lowering operators, respectively, 
\begin{equation}
 \tilde A_1^\dagger \vert n_1,n_2\rangle_q = \sqrt{[n_1 +1]_q}\,\vert n_1 +1,n_2\rangle_q;\quad 
 \tilde A_2^\dagger \vert n_1,n_2\rangle_q = \sqrt{[n_2 +1]_q}\,\vert n_1,n_2 + 1\rangle_q,
\end{equation}
and 
\begin{equation}
 \tilde A_1\vert n_1,n_2\rangle_q = \sqrt{[n_1]_q}\,\vert n_1-1,n_2\rangle_q;\quad 
 \tilde A_2\vert n_1,n_2\rangle_q = \sqrt{[n_2]_q}\,\vert n_1,n_2-1\rangle_q \,.
\end{equation}
Using these states, we present the q-deformed bi-coherent states for the Hamiltonian $\tilde H_q$ as follows
\begin{equation}\label{gcoh}
 \vert J_1,\gamma_1;\; J_2,\gamma_2\rangle_q = \frac{1}{\sqrt{E_q(J_1,J_2)}}\sum_{n_1,n_2 = 0}^\infty
 \frac{J_1^{\frac{n_1}{2}}J_2^{\frac{n_2}{2}}\exp{-i(\gamma_1 [n_1]_q + 
 \gamma_2 [n_2]_q)}}{\sqrt{[n_1]_q![n_2]_q!}}\vert n_1,n_2\rangle_q ,
\end{equation}
where 
\begin{equation}
  E_q(J_1,J_2)= 
 \sum_{n_1,n_2 = 0}^{\infty}\frac{J_1^{n_1}J_2^{n_2}}{[n_1]_q![n_2]_q!} = \mathcal{N}^2(J_1,J_2);\; 
 0 \le J_i < \frac{1}{1-q^2}, i =1,2 .
\end{equation}

\section{Generalized Heisenberg's uncertainty relations}\label{sec3}

 In order to verify the uncertainty relations, we need first to set the 
 canonical variables corresponding to the algebra (\ref{daacc}). It is 
 convenient to use the following setting
 \begin{eqnarray}\nonumber 
 \tilde A_1 &=& \frac{1}{\sqrt{K_1}}
\left( -\frac{\lambda_1}{\hbar}\hat X_{1,q} - i\hat P_{1,q} -
i\frac{\lambda_1}{\hbar}\hat X_{2,q} +\hat P_{2,q} \right);\;
  \tilde A_1^\dagger = \frac{1}{\sqrt{K_1}}
\left( -\frac{\lambda_1}{\hbar}\hat X_{1,q} + i\hat P_{1,q}
+i\frac{\lambda_1}{\hbar}\hat X_{2,q} +\hat P_{2,q} \right);\\\label{qca}
 \tilde A_2 &=& \frac{1}{\sqrt {K_2}}
\left( \frac{\lambda_2}{\hbar}\hat X_{1,q} + i\hat P_{1,q}
-i\frac{\lambda_2}{\hbar}\hat X_{2,q} +\hat P_{2,q} \right);\;
 \tilde A_2^\dagger = \frac{1}{\sqrt{K_2}}
\left( \frac{\lambda_2}{\hbar}\hat X_{1,q} - i\hat P_{1,q} +
i\frac{\lambda_2}{\hbar}\hat X_{2,q} +\hat P_{2,q} \right),
\end{eqnarray}
 where $\hat X_{i,q},\;\hat P_{i,q}, \; i=1,2$ are the corresponding 
 canonical variables. Inverting the equations in (\ref{qca}), we have
 \begin{equation}
 \hat X_{1,q} =  \frac{-\hbar\sqrt{K_1}}{2(\lambda_1 +\lambda_2)}\left(\tilde A_1 + \tilde A_1^\dagger\right)
+ \frac{\hbar\sqrt{K_2}}{2(\lambda_1 +\lambda_2)}\left(\tilde A_2 + \tilde A_2^\dagger\right);
\end{equation}
\begin{equation}
 \hat X_{2,q} = \frac{i\hbar\sqrt{K_1}}{2(\lambda_1 +\lambda_2)}\left(\tilde A_1 -\tilde A_1^\dagger\right)
 +\frac{i\hbar\sqrt{K_2}}{2(\lambda_1 +\lambda_2)}\left(\tilde A_2 -\tilde A^\dagger_2\right);
 \end{equation}
 \begin{equation}
 \hat P_{1,q} = \frac{i\lambda_2\sqrt{K_1}}{2(\lambda_1 +\lambda_2)}\left(\tilde A_1 -\tilde A_1^\dagger\right)
 -\frac{i\lambda_1\sqrt{K_2}}{2(\lambda_1 +\lambda_2)}\left(\tilde A_2 -\tilde A^\dagger_2\right);
 \end{equation}
 \begin{equation}
 \hat P_{2,q} = \frac{\lambda_2\sqrt{K_1}}{2(\lambda_1 +\lambda_2)}\left(\tilde A_1 + \tilde A_1^\dagger\right) 
+ \frac{\lambda_1\sqrt{K_2}}{2(\lambda_1 +\lambda_2)}\left(\tilde A_2 + \tilde A_2^\dagger\right),
 \end{equation}
 and they satisfy the deformed canonical commutation relations
 \begin{equation}\label{qc1}
  \left[\hat X_{1,q},\; \hat X_{2,q}\right] = i\theta + (1-q^2)\left(\frac{i\hbar^2}{2(\lambda_1 +\lambda_2)^2}\right)
  \left(K_2 \tilde A_2^\dagger \tilde A_2 - K_1\tilde A_1^\dagger \tilde A_1\right);
  \end{equation}
  \begin{equation}\label{qc2}
  \left[ \hat X_{1,q},\;\hat P_{1,q}\right] = i\hbar - (1-q^2)\left(\frac{i\hbar}{2(\lambda_1 +\lambda_2)^2}\right)
  \left( \lambda_1 K_2 \tilde A_2^\dagger \tilde A_2 + \lambda_2 K_1 \tilde A_1^\dagger \tilde A_1 \right);
  \end{equation}
  \begin{equation}\label{qc3}
  \left[ \hat X_{2,q},\;\hat P_{2,q}\right] = i\hbar - (1-q^2)\left(\frac{i\hbar}{2(\lambda_1 +\lambda_2)^2}\right)
  \left(\lambda_1 K_2 \tilde A_2^\dagger \tilde A_2 + \lambda_2 K_1 \tilde A_1^\dagger \tilde A_1 \right);
  \end{equation}
  \begin{eqnarray}\label{qc4}
  \left[\hat P_{1,q},\;\hat P_{2,q}\right] &=& i(1-q^2)\left(\frac{1}{2(\lambda_1 +\lambda_2)^2}\right)
  \left(\lambda_1^2K_2 \tilde A_2^\dagger \tilde A_2 -\lambda_2^2 K_1 \tilde A_1^\dagger \tilde A_1\right) ; \\
  \left[\hat X_{1,q},\;\hat P_{2,q}\right] &=& 0 ; \\
  \left[\hat X_{2,q},\;\hat P_{1,q}\right] &=& 0.
  \end{eqnarray}

Next we use the expressions in equation (\ref{qca}) and evaluate
\begin{eqnarray}\nonumber
 K_1 \tilde A_1^\dagger \tilde A_1 &=& \left(\frac{\lambda_1}{\hbar}\right)^2 \hat X_{1,q}^2 +
 \left(\frac{\lambda_1}{\hbar}\right)^2 \hat X_{2,q}^2 + \hat P_{1,q}^2 + \hat P_{2,q}^2 
 -\frac{2\lambda_1}{\hbar}\hat X_{1,q} \hat P_{2,q} + \frac{2\lambda_1}{\hbar}\hat X_{2,q}\hat P_{1,q}\\\label{k1aa}
 &+& i \left(\frac{\lambda_1}{\hbar}\right)^2 [\hat X_{1,q},\hat X_{2,q}] + 
 i\frac{\lambda_1}{\hbar}[\hat X_{1,q},\hat P_{1,q}] +i\frac{\lambda_1}{\hbar}[\hat X_{2,q},\hat P_{2,q}] 
  +i [\hat P_{1,q},\hat P_{2,q}];
\end{eqnarray}
\begin{eqnarray}\nonumber
 K_2 \tilde A_2^\dagger \tilde A_2 &=& \left(\frac{\lambda_2}{\hbar}\right)^2 \hat X_{1,q}^2 +
 \left(\frac{\lambda_2}{\hbar}\right)^2 \hat X_{2,q}^2 + \hat P_{1,q}^2 + \hat P_{2,q}^2 
 + \frac{2\lambda_2}{\hbar}\hat X_{1,q} \hat P_{2,q} - \frac{2\lambda_2}{\hbar}\hat X_{2,q}\hat P_{1,q}\\\label{k2aa}
 &-& i \left(\frac{\lambda_2}{\hbar}\right)^2 [\hat X_{1,q},\hat X_{2,q}] + 
 i\frac{\lambda_2}{\hbar}[\hat X_{1,q},\hat P_{1,q}] +i\frac{\lambda_2}{\hbar}[\hat X_{2,q},\hat P_{2,q}] 
  -i [\hat P_{1,q},\hat P_{2,q}].
\end{eqnarray}
Substituting (\ref{k1aa}) and (\ref{k2aa}) into the right-hand side of (\ref{qc1})-(\ref{qc4}) we obtain four 
equations for the four unknown commutators $[\hat X_{1,q},\hat X_{2,q}],\,[\hat X_{1,q},\hat P_{1,q}],\,
[\hat X_{2,q},\hat P_{2,q}],\, [\hat P_{1,q},\hat P_{2,q}]$. Solving these equations, we obtain the following
noncommutative relations:
\begin{eqnarray}\nonumber
 \left[\hat X_{1,q},\hat X_{2,q}\right] = \frac{2 i \theta}{1 + q^2}
 &+& i\frac{1-q^2}{1+q^2}\left( \frac{\lambda_2-\lambda_1}{\lambda_1 +\lambda_2}
 \hat X_{1,q}^2 + \frac{\lambda_2-\lambda_1}{\lambda_1 +\lambda_2}\hat X_{2,q}^2\right.\\\label{q1aa}
 &+& \left. 2\hbar\frac{\left((\lambda_1 +\lambda_2)^3 -(1-q^2)(\lambda_1 +\lambda_2 -1)
 \lambda_1\lambda_2\right)}{(\lambda_1 +\lambda_2)^4}(\hat X_{1,q}\hat P_{2,q}
 -\hat X_{2,q}\hat P_{1,q})\right);
\end{eqnarray}
\begin{eqnarray}\nonumber
 \left[\hat X_{1,q},\hat P_{1,q}\right] = \frac{2 i \hbar}{1 + q^2}
 &-& i\frac{1-q^2}{1+q^2}\left( \frac{\lambda_1\lambda_2}{\lambda_1 +\lambda_2}\hat X_{1,q}^2 + 
 \frac{\lambda_1\lambda_2}{\lambda_1 +\lambda_2}\hat X_{2,q}^2\right.\\\nonumber
 &+& \left. \frac{2\hbar}{\lambda_1 +\lambda_2}\hat P_{1,q}^2 + \frac{2\hbar}{\lambda_1 +\lambda_2}\hat P_{2,q}^2\right.\\\label{q2aa}
 &-& \left. 2\frac{(1-q^2)(\lambda_1 -\lambda_2)(\lambda_1 +\lambda_2 -1)\lambda_1\lambda_2}
 {(\lambda_1 +\lambda_2)^4}(\hat X_{1,q}\hat P_{2,q}
 -\hat X_{2,q}\hat P_{1,q})\right);
\end{eqnarray}
\begin{eqnarray}\nonumber
 \left[\hat X_{2,q},\hat P_{2,q}\right] = \frac{2 i \hbar}{1 + q^2}
 &-& i\frac{1-q^2}{1+q^2}\left( \frac{\lambda_1\lambda_2}{\lambda_1 +\lambda_2}\hat X_{1,q}^2 + 
 \frac{\lambda_1\lambda_2}{\lambda_1 +\lambda_2}\hat X_{2,q}^2\right.\\\nonumber
 &+& \left. \frac{2\hbar}{\lambda_1 +\lambda_2}\hat P_{1,q}^2 + \frac{2\hbar}{\lambda_1 +\lambda_2}\hat P_{2,q}^2\right.\\\label{q3aa}
 &-& \left. 2\frac{(1-q^2)(\lambda_1 -\lambda_2)(\lambda_1 +\lambda_2 -1)\lambda_1\lambda_2}
 {(\lambda_1 +\lambda_2)^4}(\hat X_{1,q}\hat P_{2,q}
 -\hat X_{2,q}\hat P_{1,q})\right);
\end{eqnarray}
\begin{eqnarray}\nonumber
\left[\hat P_{1,q},\hat P_{2,q}\right] &= &
i\frac{1-q^2}{1+q^2}\left( \frac{(\hbar -1)(\lambda_1 -\lambda_2)\lambda_1\lambda_2}{\hbar^2(\lambda_1 +\lambda_2)}\hat X_{1,q}^2 + 
\frac{(\hbar -1)(\lambda_1 -\lambda_2)\lambda_1\lambda_2}{\hbar^2(\lambda_1 +\lambda_2)} 
\hat X_{2,q}^2\right.\\\nonumber
 &+& \left. \frac{2(\lambda_1-\lambda_2)}{\lambda_1 +\lambda_2}\hat P_{1,q}^2 + 
\frac{2(\lambda_1-\lambda_2)}{\lambda_1 +\lambda_2}\hat P_{2,q}^2\right.\\\nonumber
&+& \left. \lambda_1\lambda_2\left[\frac{2(\lambda_1 +\lambda_2)^3 -(1+q^2)(\lambda_1 +\lambda_2 -1)
(\lambda_1 +\lambda_2)^2 }{\hbar(\lambda_1 +\lambda_2)^4}\right.\right.\\\label{q4aa}
&-&\left.\left. \frac{2(1-q^2)(\lambda_1 +\lambda_2 -1)(\lambda_1^2 +\lambda_2^2 -\lambda_1\lambda_2))}
{\hbar(\lambda_1 +\lambda_2)^4}\right](\hat X_{1,q}\hat P_{2,q}
-\hat X_{2,q}\hat P_{1,q})\right).
\end{eqnarray}
The commutators $\left[\hat X_{2,q},\;\hat P_{1,q}\right]$ and $\left[\hat X_{1,q},\;\hat P_{2,q}\right]$ remain zero.
At the limit $ q \to 1$, we recover the algebras in equations (\ref{NCHA}) and (\ref{aacc}). Note that if $\theta =0 $
we have $\lambda_1 =\lambda_2 = \hbar m\omega$ and $K_1 = K_2 = 4 \hbar m \omega $.

The commutation relations in equations (\ref{q1aa}), (\ref{q2aa}), (\ref{q3aa}) 
and (\ref{q4aa}) are dynamical since they are expressed in terms of 
the dynamical variables that are $\hat X_{1,q},\hat X_{2,q},\hat P_{1,q},\hat P_{2,q}$, by dynamical variables we mean observables. 
This is a consequence of the q-deformation since 
at the limit $\theta \to 0 $, they remain dynamical. 
The generalized Heisenberg uncertainty relation for a simultaneous measurement of two observables 
$\mathcal{O}_1$ and $\mathcal{O}_2$ , 
where $\mathcal{O}_1, \mathcal{O}_2\in \{\hat X_{i,q}, \hat P_{i,q},\; i=1,2\}$, projected onto the normalized coherent states 
$\vert J_1,\gamma_1; J_2,\gamma_2\rangle_q $ defined in equation (\ref{gcoh}) 
\begin{eqnarray}\label{gur}
 \Delta \mathcal{O}_1\Delta \mathcal{O}_2\vert_{\vert J_1,\gamma_1; J_2,\gamma_2\rangle_q}\ge \frac{1}{2}\vert 
 \left({}_q\langle J_1,\gamma_1; J_2,\gamma_2\vert \left[\mathcal{O}_1,\mathcal{O}_2\right]
 \vert J_1,\gamma_1; J_2,\gamma_2\rangle_q\right)\vert,
\end{eqnarray}
where the uncertainty for $\mathcal{O}$, $\mathcal{O} \in \{\hat X_{i,q},\;\hat P_{i,q},\; i=1,2\}$ is computed as 
\begin{equation}\label{uncert}
\Delta \mathcal{O}^2 = {}_q\langle J_1,\gamma_1; J_2,
\gamma_2\vert \mathcal{O}^2\vert J_1,\gamma_1; J_2,\gamma_2\rangle_q  -
\left({}_q\langle J_1,\gamma_1; J_2,\gamma_2\vert \mathcal{O} \vert J_1,\gamma_1; J_2,\gamma_2\rangle_q\right)^2.
\end{equation}
We start by evaluating the left-hand side of equations (\ref{gur}), where the 
details of the calculations are given in the Appendix in section \ref{app}.
\begin{eqnarray}\label{d1}
 \Delta \hat X_{1,q}^2\Delta \hat X_{2,q}^2 
 &=& \chi_1(J_1,J_2;\gamma_1,\gamma_2)\; \chi_2(J_1,J_2;\gamma_1,\gamma_2);\\\label{d2}
 \Delta \hat X_{1,q}^2\Delta \hat P_{1,q}^2 
 &=& \chi_1(J_1,J_2;\gamma_1,\gamma_2)\;\kappa_1(J_1,J_2,\gamma_1,\gamma_2);\\\label{d3}
 \Delta \hat X_{2,q}^2\Delta \hat P_{2,q}^2 
 &=& \chi^2(J_1,J_2,\gamma_1,\gamma_2)\;\kappa_2(J_1,J_2;\gamma_1,\gamma_2);\\\label{d4}
 \Delta \hat P_{1,q}^2\Delta \hat P_{2,q}^2 
 &=& \kappa_1(J_1,J_1;\gamma_1,\gamma_2)\;\kappa_2(J_1,J_2;\gamma_1,\gamma_2),
\end{eqnarray}
where 
 \begin{eqnarray}\nonumber
 \chi_1 (J_1,J_2;\gamma_1,\gamma_2) := \Delta \hat X_{1,q}^2  &=&    \frac{\hbar^2K_1}{4(\lambda_1+\lambda_2)^2}
 \left(1 + (1+q^2)J_1 + G_q(\gamma_1) - G_c^2(\gamma_1)\right)\\\nonumber
  &+& \frac{\hbar^2K_2}{4(\lambda_1+\lambda_2)^2}
 \left(1 + (1+q^2)J_2 + G_q(\gamma_2) - G_c^2(\gamma_2)\right)\\
 &+& \frac{\hbar^2\sqrt{K_1K_2}}{2(\lambda_1+\lambda_2)^2}
 \left( -G^+_c(\gamma_1,\gamma_2) - G_c^-(\gamma_1,\gamma_2) +
 G_c(\gamma_1)G_c(\gamma_2)\right);
 \end{eqnarray}
\begin{eqnarray}\nonumber
 \chi_2(J_1,J_2;\gamma_1,\gamma_2) := \Delta \hat X_{2,q}^2 &=& \frac{\hbar^2K_1}{4(\lambda_1+\lambda_2)^2}
 \left(1 + (1+q^2)J_1 - G_q(\gamma_1) + G_s^2(\gamma_1)\right)\\\nonumber
 &+& \frac{\hbar^2K_2}{4(\lambda_1+\lambda_2)^2}
 \left(1 + (1+q^2)J_2 - G_q(\gamma_2) + G_s^2(\gamma_2)\right)\\
 &+& \frac{\hbar^2\sqrt{K_1K_2}}{2(\lambda_1+\lambda_2)^2}
 \left( -G^+_c(\gamma_1,\gamma_2) + G_c^-(\gamma_1,\gamma_2) +
 G_s(\gamma_1)G_s(\gamma_2)\right);
\end{eqnarray}
\begin{eqnarray}\nonumber
\kappa_1(J_1,J_2;\gamma_1,\gamma_2) := \Delta \hat P_{1,q}^2 &=&  \frac{\lambda_2^2K_1}{4(\lambda_1+\lambda_2)^2}
 \left(1 + (1+q^2)J_1 - G_q(\gamma_1) + G_s^2(\gamma_1)\right)\\\nonumber
 &+& \frac{\lambda_1^2K_2}{4(\lambda_1+\lambda_2)^2}
 \left(1 + (1+q^2)J_2 - G_q(\gamma_2) + G_s^2(\gamma_2)\right)\\
 &+& \frac{\lambda_1\lambda_2\sqrt{K_1K_2}}{2(\lambda_1+\lambda_2)^2}
 \left( G^+_c(\gamma_1,\gamma_2) - G_c^-(\gamma_1,\gamma_2) 
 -G_s(\gamma_1)G_s(\gamma_2)\right);
\end{eqnarray}
\begin{eqnarray}\nonumber
\kappa_2(J_1,J_2;\gamma_1,\gamma_2) := \Delta \hat P_{2,q}^2 &=& 
  \frac{\lambda_2^2K_1}{4(\lambda_1+\lambda_2)^2}
 \left(1 + (1+q^2)J_1 + G_q(\gamma_1) - G_c^2(\gamma_1)\right)\\\nonumber
 &+& \frac{\lambda_1^2K_2}{4(\lambda_1+\lambda_2)^2}
 \left(1 + (1+q^2)J_2 + G_q(\gamma_2) - G_c^2(\gamma_2)\right)\\
 &+& \frac{\lambda_1\lambda_2\sqrt{K_1K_2}}{2(\lambda_1+\lambda_2)^2}
 \left( G^+_c(\gamma_1,\gamma_2) + G_c^-(\gamma_1,\gamma_2) 
 - G_c(\gamma_1)G_c(\gamma_2)\right).
\end{eqnarray}

Using the commutation relations (\ref{qc1}), (\ref{qc2}), (\ref{qc3}), (\ref{qc4}) 
and the equations (\ref{ee1}) and (\ref{ee2}) in the Appendix (see in \ref{app} ), we evaluate the  
the corresponding expressions in the right-hand side of equation (\ref{gur}) as follows
 \begin{equation}\label{heis1}
 \frac{1}{2}\vert 
 \left({}_q\langle J_1,\gamma_1; J_2,\gamma_2\vert \left[\hat X_{1,q},\hat X_{2,q}\right]
 \vert J_1,\gamma_1; J_2,\gamma_2\rangle_q\right)\vert = 
 \frac{\hbar^2}{4(\lambda_1 +\lambda_2)^2}\vert K_1\left(1-(1-q^2)J_1\right)-
 K_2\left(1 -(1-q^2)J_2\right)\vert;
 \end{equation}
 \begin{equation}\label{heis2}
 \frac{1}{2}\vert 
 \left({}_q\langle J_1,\gamma_1; J_2,\gamma_2\vert \left[\hat X_{1,q},\hat P_{1,q}\right]
 \vert J_1,\gamma_1; J_2,\gamma_2\rangle_q\right)\vert =
 \frac{\hbar}{4(\lambda_1 +\lambda_2)^2}\vert 
 \lambda_2K_1\left(1-(1-q^2)J_1\right) +
 \lambda_1K_2\left(1 -(1-q^2)J_2\right)\vert;
 \end{equation}
 \begin{equation}\label{heis3}
 \frac{1}{2}\vert 
 \left({}_q\langle J_1,\gamma_1; J_2,\gamma_2\vert \left[\hat X_{2,q},\hat P_{2,q}\right]
 \vert J_1,\gamma_1; J_2,\gamma_2\rangle_q\right)\vert =
 \frac{\hbar}{4(\lambda_1 +\lambda_2)^2}\vert 
 \lambda_2K_1\left(1-(1-q^2)J_1\right) +
 \lambda_1K_2\left(1 -(1-q^2)J_2\right)\vert;
 \end{equation}
 \begin{equation}\label{heis4}
 \frac{1}{2}\vert 
 \left({}_q\langle J_1,\gamma_1; J_2,\gamma_2\vert \left[\hat P_{1,q},\hat P_{2,q}\right]
 \vert J_1,\gamma_1; J_2,\gamma_2\rangle_q\right)\vert =
 \frac{1}{4(\lambda_1 +\lambda_2)^2}\vert \lambda_2^2K_1(1 -(1-q^2)J_1) - \lambda_1^2 K_2(1 -(1-q^2)J_2)\vert.
\end{equation}
A first remark is that the right hand side of the generalized Heisenberg's inequality (\ref{gur}), that are 
respectively the equations (\ref{heis1}),(\ref{heis2}), (\ref{heis3}), (\ref{heis4}), are constant value independent 
of $\gamma_i, \; i = 1,2$. So noting the following limits 
$\lim_{\gamma_1\to 0}G_c(\gamma_1) = 2\sqrt{J_1}$,
 $\lim_{\gamma_1\to 0}G_s(\gamma_1)= 0$,
 $\lim_{\gamma_1\to 0}G_q(\gamma_1)= 2 J_1$,
 $\lim_{\gamma_2\to 0}G_c(\gamma_2) = 2\sqrt{J_2}$,
 $\lim_{\gamma_2\to 0}G_s(\gamma_2)= 0$,
 $\lim_{\gamma_2\to 0}G_q(\gamma_2)= 2 J_2$ and $\lim_{\gamma_1\to 0,\gamma_2\to 0}G^+_c(\gamma_1,\gamma_2) = 2\sqrt{J_1J_2}$,
 $\lim_{\gamma_1\to 0,\gamma_2\to 0}G^+_s(\gamma_1,\gamma_2) = 0 $,
 $\lim_{\gamma_1\to 0,\gamma_2\to 0}G^-_c(\gamma_1,\gamma_2) = 2\sqrt{J_1J_2}$,
 $\lim_{\gamma_1\to 0,\gamma_2\to 0}G^-_s(\gamma_1,\gamma_2) = 0$,
we see that when  $\gamma_1 = \gamma_2 = 0$, the equations (\ref{d1}),(\ref{d2})
(\ref{d3}), (\ref{d4}) are respectively the following 
\begin{eqnarray}\nonumber
 \Delta \hat X_{1,q}^2\Delta \hat X_{2,q}^2 &=& \frac{\hbar^2}{4(\lambda_1 +\lambda_2)^2}
 \left(K_1(1 -(1-q^2)J_1) + K_2(1 -(1-q^2)J_2)\right)\\
 &\times & \frac{\hbar^2}{4(\lambda_1 +\lambda_2)^2}
 \left(K_1(1 -(1-q^2)J_1) + K_2(1 -(1-q^2)J_2)\right);
 \end{eqnarray}
 \begin{eqnarray}\nonumber
 \Delta \hat X_{1,q}^2\Delta \hat P_{1,q}^2 &=& \left(\frac{\hbar}{4(\lambda_1 +\lambda_2)^2}\right)^2
 \left[\lambda_2K_1(1 -(1-q^2)J_1) + \lambda_1K_2(1 -(1-q^2)J_2)\right]^2\\\label{sat1}
 &+& \left(\frac{\hbar (\lambda_1 -\lambda_2)}{4(\lambda_1 +\lambda_2)^2}\right)^2
 K_1K_2\left(1 -(1-q^2)J_1\right)\left(1 -(1-q^2)J_2)\right);
 \end{eqnarray}
 \begin{eqnarray}\nonumber
 \Delta \hat X_{2,q}^2\Delta \hat P_{2,q}^2 &=&
 \left(\frac{\hbar}{4(\lambda_1 +\lambda_2)^2}\right)^2
 \left[\lambda_2K_1(1 -(1-q^2)J_1) + \lambda_1K_2(1 -(1-q^2)J_2)\right]^2\\\label{sat2}
 &+& \left(\frac{\hbar (\lambda_1 -\lambda_2)}{4(\lambda_1 +\lambda_2)^2}\right)^2
 K_1K_2\left(1 -(1-q^2)J_1\right)\left(1 -(1-q^2)J_2)\right);
 \end{eqnarray}
 \begin{eqnarray}\nonumber
 \Delta \hat P_{1,q}^2\Delta \hat P_{2,q}^2 &=& 
 \frac{1}{4(\lambda_1 +\lambda_2)^2}
 \left(\lambda_2^2K_1(1 -(1-q^2)J_1) + \lambda_1^2 K_2(1 -(1-q^2)J_2)\right)\\
 &\times &
 \frac{1}{4(\lambda_1 +\lambda_2)^2}
 \left(\lambda_2^2K_1(1 -(1-q^2)J_1) + \lambda_1^2 K_2(1 -(1-q^2)J_2)\right).
\end{eqnarray}
Let us analyse these results. For $\gamma_i = 0, i =1,2 $, 
the inequality (\ref{gur}) is always respected. When in addition  
the parameter of noncommutativity $\theta = 0$, where $\lambda_1 =\lambda_2 =\hbar m\omega$ 
and $K_1 =K_2 =4\hbar m\omega$, the expressions (\ref{sat1}) 
and (\ref{sat2}) become respectively the square of the expressions (\ref{heis2})
and (\ref{heis3}) such that the inequality is saturated  for a simultaneous 
measurement of the position -momentum  observables.
Let us discuss now the general case where $\gamma_1\neq 0$ and $\gamma_2\neq 0$. In 
that case, comparing respectively the equations (\ref{d1} -\ref{d4}) 
and (\ref{heis1}-\ref{heis4}), we require some conditions, that are
the expressions (\ref{p1}), (\ref{p2}), (\ref{p3}) and (\ref{p4}) in 
the Appendix, in order for the inequality (\ref{gur}) 
to be satisfied. Reducing the inequalities (\ref{p1}-\ref{p2}) we require the conditions: 
$4J_i-G_c^2(\gamma_i) + G_s^2(\gamma_i)\ge 0,\;i=1,2$ and 
$2G^+_c(\gamma_1,\gamma_2) -\left(G_c(\gamma_1)G_c(\gamma_2) + G_s(\gamma_1)G_s(\gamma_2)\right)\ge 0$.
Assuming that $\lambda_1\lambda_2 -1\ge 0$  that means 
$\hbar^2 m^2\omega^2 - 1\ge 0$ and noting the range of the functions $G_c(\gamma_i), G_s(\gamma_i),\; i =1,2; $ 
and $G_c^+(\gamma_1,\gamma_2)$
are respectively $-2\sqrt{J_i}\le G_c(\gamma_i) \le 2\sqrt{J_i},\; -2i\sqrt{J_i}\le G_s(\gamma_i) \le 2i\sqrt{J_i},
\;i= 1,2$ and \- $-2\sqrt{J_1J_2} \le G_c^+(\gamma_1,\gamma_2)\le 2\sqrt{J_1J_2}$, the uncertainty relations (\ref{gur})
are not always satisfied.

\section{Conclusion}\label{sec4}

The analysis in \cite{fring} has been extended to the case of the two dimensional
non-commutative harmonic oscillator where we constructed time dependent 
q-deformed bi-coherent states. Since q-deformed oscillator algebras have 
been shown to be related to noncommutative space-time structures, 
the presence of $\theta$ may be considered unnecessary. As we have shown, 
for $\gamma_i = 0,\; i = 1,2$ and setting $\theta = 0$ occurs the saturation of the inequality for a 
simultaneous measurement of the position and momentum observables and that is not valid in
the presence of the parameter of noncommutativity $\theta$. 
For the general case, where $\gamma_i \neq 0,\; i =1,2$, we found that the generalized uncertainty 
relations (\ref{gur}) are not always respected. That is not the case in \cite{fring}.
An extension of this work would be to study the revival time structures, 
it may also occur minimal lengths and minimal momenta from the commutation relations 
(\ref{q1aa} - \ref{q4aa}) to be computed.  The 
fermionic coherent states have been recently discussed in \cite{monic} and 
previously the q-deformed fermions have also been studied \cite{nara}, \cite{nathan}.
An interesting extension of this work would be to consider q-deformed fermionic operators 
for coherent states and compare with the bosonic case.

\section{Appendix: Details of calculations of the equations in section (\ref{sec3})}\label{app}

Let's compute the square of the canonical operators since they appears in the calculations of the uncertainties
\begin{eqnarray}\nonumber
 \hat X_{1,q}^2 &=& \frac{\hbar^2K_1}{4(\lambda_1 +\lambda_2)^2}\left(\tilde A_1\tilde A_1 + \tilde A_1\tilde A_1^\dagger 
 + \tilde A_1^\dagger \tilde A_1 + \tilde A_1^\dagger \tilde A_1^\dagger\right) 
 + \frac{\hbar^2K_2}{4(\lambda_1 +\lambda_2)^2}\left( \tilde A_2\tilde A_2 + \tilde A_2\tilde A_2^\dagger 
 + \tilde A_2^\dagger \tilde A_2 + \tilde A_2^\dagger \tilde A_2^\dagger \right)\\
 &-& \frac{\hbar^2\sqrt{K_1K_2}}{2(\lambda_1 + \lambda_2)^2}\left(\tilde A_1\tilde A_2 + \tilde A_1\tilde A_2^\dagger 
 + \tilde A_1^\dagger \tilde A_2 + \tilde A_1^\dagger \tilde A_2^\dagger\right);
 \end{eqnarray}
 \begin{eqnarray}\nonumber
 \hat X_{2,q}^2 &=& -\frac{\hbar^2K_1}{4(\lambda_1 +\lambda_2)^2}\left(\tilde A_1\tilde A_1 -\tilde A_1\tilde A_1^\dagger 
 -\tilde A_1^\dagger \tilde A_1 + \tilde A_1^\dagger \tilde A_1^\dagger\right) 
 -\frac{\hbar^2 K_2}{4(\lambda_1 +\lambda_2)^2}\left(\tilde A_2\tilde A_2 -\tilde A_2^\dagger \tilde A_2 -\tilde A_2 \tilde A_2^\dagger 
 + \tilde A_2^\dagger \tilde A_2^\dagger \right)\\
 &-& \frac{\hbar^2\sqrt{K_1K_2}}{2(\lambda_1 +\lambda_2)^2}\left(\tilde A_1\tilde A_2 -\tilde A_1\tilde A_2^\dagger -
 \tilde A_1^\dagger \tilde A_2 + \tilde A_1^\dagger \tilde A_2^\dagger\right); 
 \end{eqnarray}
 \begin{eqnarray}\nonumber
 \hat P_{1,q}^2 &=& -\frac{\lambda_2^2K_1}{4(\lambda_1 +\lambda_2)^2}\left(\tilde A_1\tilde A_1 -\tilde A_1^\dagger \tilde A_1 
 -\tilde A_1\tilde A_1^\dagger + \tilde A_1^\dagger \tilde A_1^\dagger\right) 
 -\frac{\lambda_1^2K_2}{4(\lambda_1 +\lambda_2)^2}\left(\tilde A_2\tilde A_2 -\tilde A_2^\dagger \tilde A_2 - 
 \tilde A_2\tilde A_2^\dagger + \tilde A_2^\dagger \tilde A_2^\dagger\right)\\
 &+&\frac{\lambda_1\lambda_2\sqrt{K_1K_2}}{2(\lambda_1 +\lambda_2)^2}\left(\tilde A_1\tilde A_2 -\tilde A_1\tilde A_2^\dagger 
 -\tilde A_1^\dagger \tilde A_2 + \tilde A_1^\dagger \tilde A_2^\dagger\right);
\end{eqnarray}
\begin{eqnarray}\nonumber
 \hat P_{2,q}^2 &=& \frac{\lambda_2^2K_1}{4(\lambda_1 +\lambda_2)^2}\left(\tilde A_1\tilde A_1 + \tilde A_1\tilde A_1^\dagger 
 + \tilde A_1^\dagger \tilde A_1 + \tilde A_1^\dagger \tilde A_1^\dagger\right) 
 + \frac{\lambda_1^2K_2}{4(\lambda_1 +\lambda_2)^2}\left( \tilde A_2\tilde A_2 + \tilde A_2\tilde A_2^\dagger + 
 \tilde A_2^\dagger \tilde A_2 + \tilde A_2^\dagger \tilde A_2^\dagger \right)\\
 &+&\frac{\lambda_1\lambda_2\sqrt{K_1K_2}}{2(\lambda_1 +\lambda_2)^2}\left(\tilde A_1\tilde A_2 + \tilde A_1\tilde A_2^\dagger + 
 \tilde A_1^\dagger \tilde A_2 + \tilde A_1^\dagger \tilde A_2^\dagger\right).
\end{eqnarray}
In order to verify the inequality (\ref{gur}) for the states (\ref{gcoh}), we compute first the 
expectation values for the creation and annihilation operators $A_i, \; A_i^\dagger, i=1,2$
\begin{equation}
{}_q\langle J_1,\gamma_1; J_2,\gamma_2\vert \tilde A_1 \vert J_1,\gamma_1; J_2,\gamma_2\rangle_q = 
\frac{\sqrt{J_1}}{E_q(J_1,J_2)}F_q(J_1,J_2,-\gamma_1);
\end{equation}
\begin{equation}
{}_q\langle J_1,\gamma_1; J_2,\gamma_2\vert \tilde A_2 \vert J_1,\gamma_1; J_2,\gamma_2\rangle_q =
\frac{\sqrt{J_2}}{E_q(J_1,J_2)}F_q(J_1,J_2,-\gamma_2);
\end{equation}
\begin{equation}
{}_q\langle J_1,\gamma_1; J_2,\gamma_2\vert \tilde A_1^\dagger \vert J_1,\gamma_1; J_2,\gamma_2\rangle_q =
\frac{\sqrt{J_1}}{E_q(J_1,J_2)}F_q(J_1,J_2,\gamma_1);
\end{equation}
\begin{equation}
{}_q\langle J_1,\gamma_1; J_2,\gamma_2\vert \tilde A_2^\dagger \vert J_1,\gamma_1; J_2,\gamma_2\rangle_q =
\frac{\sqrt{J_2}}{E_q(J_1,J_2)}F_q(J_1,J_2,\gamma_2),
\end{equation}
where we introduced the function
\begin{equation}
 F_q(J_1,J_2,\gamma_1) := \sum_{n_1,n_2 =0}^\infty \frac{J_1^{n_1}J_2^{n_2}e^{i\gamma_1 q^{2 [n_1]_q}}}{[n_1]_q![n_2]_q!};\quad 
 F_q(J_1,J_2,\gamma_2) := \sum_{n_1,n_2 =0}^\infty \frac{J_1^{n_1}J_2^{n_2}e^{i\gamma_2 q^{2 [n_2]_q}}}{[n_1]_q![n_2]_q!}
\end{equation}
In order to compute the expectation values for $\mathcal{O}_i^2$, $\mathcal{O}\in \{\hat X_i,\hat P_i,\; i=1,2  \}$, we use the 
their expressions as in the appendix. We evaluate then
\begin{eqnarray}
 {}_q\langle J_1,\gamma_1; J_2,\gamma_2\vert \tilde A_1\tilde A_1 \vert J_1,\gamma_1; J_2,\gamma_2\rangle_q &=& 
 \frac{J_1}{E_q(J_1,J_2)}F_q(J_1,J_2, -(1+q^2)\gamma_1)\\
 {}_q\langle J_1,\gamma_1; J_2,\gamma_2\vert \tilde A_1^\dagger \tilde A_1^\dagger \vert J_1,\gamma_1; J_2,\gamma_2\rangle_q &=&
 \frac{J_1}{E_q(J_1,J_2)}F_q(J_1,J_2, (1+q^2)\gamma_1)\\\label{ee1}
 {}_q\langle J_1,\gamma_1; J_2,\gamma_2\vert \tilde A_1^\dagger \tilde A_1 \vert J_1,\gamma_1; J_2,\gamma_2\rangle_q &=& J_1\\
 {}_q\langle J_1,\gamma_1; J_2,\gamma_2\vert \tilde A_1 \tilde A_1^\dagger \vert J_1,\gamma_1; J_2,\gamma_2\rangle_q &=& 1+ q^2J_1.
\end{eqnarray}
\begin{eqnarray}
 {}_q\langle J_1,\gamma_1; J_2,\gamma_2\vert \tilde A_2\tilde A_2 \vert J_1,\gamma_1; J_2,\gamma_2\rangle_q &=& 
 \frac{J_2}{E_q(J_1,J_2)}F_q(J_1,J_2, -(1+q^2)\gamma_2)\\
 {}_q\langle J_1,\gamma_1; J_2,\gamma_2\vert \tilde A_2^\dagger \tilde A_2^\dagger \vert J_1,\gamma_1; J_2,\gamma_2\rangle_q &=&
 \frac{J_2}{E_q(J_1,J_2)}F_q(J_1,J_2, (1+q^2)\gamma_2)\\\label{ee2}
 {}_q\langle J_1,\gamma_1; J_2,\gamma_2\vert \tilde A_2^\dagger  \tilde A_2 \vert J_1,\gamma_1; J_2,\gamma_2\rangle_q &=& J_2\\
 {}_q\langle J_1,\gamma_1; J_2,\gamma_2\vert \tilde A_2 \tilde A_2^\dagger \vert J_1,\gamma_1; J_2,\gamma_2\rangle_q &=& 1+ q^2J_2.
\end{eqnarray}
\begin{eqnarray}
 {}_q\langle J_1,\gamma_1; J_2,\gamma_2\vert \tilde A_1\tilde A_2 \vert J_1,\gamma_1; J_2,\gamma_2\rangle_q &=& 
 \frac{(J_1J_2)^{\frac{1}{2}}}{E_q(J_1,J_2)}F_q(J_1,J_2,-\gamma_1,-\gamma_2)\\
 {}_q\langle J_1,\gamma_1; J_2,\gamma_2\vert \tilde A_1^\dagger \tilde A_2^\dagger \vert J_1,\gamma_1; J_2,\gamma_2\rangle_q &=&
 \frac{(J_1J_2)^{\frac{1}{2}}}{E_q(J_1,J_2)}F_q(J_1,J_2,\gamma_1, \gamma_2)\\
 {}_q\langle J_1,\gamma_1; J_2,\gamma_2\vert \tilde A_2^\dagger \tilde A_1 \vert J_1,\gamma_1; J_2,\gamma_2\rangle_q &=& 
 \frac{(J_1J_2)^{\frac{1}{2}}}{E_q(J_1,J_2)}F_q(J_1,J_2,-\gamma_1,\gamma_2)\\
 {}_q\langle J_1,\gamma_1; J_2,\gamma_2\vert \tilde A_1^\dagger \tilde A_2 \vert J_1,\gamma_1; J_2,\gamma_2\rangle_q &=& 
 \frac{(J_1J_2)^{\frac{1}{2}}}{E_q(J_1,J_2)}F_q(J_1,J_2,\gamma_1,-\gamma_2)
\end{eqnarray}
where 
\begin{equation}
 F_q(J_1,J_2,\gamma_1,\gamma_2) := \sum_{n_1,n_2}\frac{J_1^{n_1}J_2^{n_2}e^{i\gamma_1q^{2 [n_1]_q}+
 i\gamma_2q^{2 [n_2]_q}}}{[n_1]_q![n_2]_q!}.
\end{equation}
The expectations values are the following
\begin{eqnarray}\nonumber
 {}_q\langle J_1,\gamma_1; J_2,\gamma_2\vert \hat X_{1,q} \vert J_1,\gamma_1; J_2,\gamma_2\rangle_q &=&
 -\frac{\hbar\sqrt{K_1}}{2(\lambda_1 +\lambda_2)}\frac{\sqrt{J_1}}{E_q(J_1,J_2)}\left(F_q(J_1,J_2,-\gamma_1)
 + F_q(J_1,J_2,\gamma_1)\right) \\
 &+& \frac{\hbar\sqrt{K_2}}{2(\lambda_1 +\lambda_2)}\frac{\sqrt{J_2}}{E_q(J_1,J_2)}
 \left(F_q(J_1,J_2,-\gamma_2)
 + F_q(J_1,J_2,\gamma_2)\right);
 \end{eqnarray}
 \begin{eqnarray}\nonumber
 {}_q\langle J_1,\gamma_1; J_2,\gamma_2\vert \hat X_{2,q} \vert J_1,\gamma_1; J_2,\gamma_2\rangle_q &=&
 \frac{i \hbar\sqrt{K_1}}{2(\lambda_1 +\lambda_2)}\frac{\sqrt{J_1}}{E_q(J_1,J_2)}\left(F_q(J_1,J_2,-\gamma_1)
 - F_q(J_1,J_2,\gamma_1)\right) \\
 &+& \frac{i \hbar\sqrt{K_2}}{2(\lambda_1 +\lambda_2)}\frac{\sqrt{J_2}}{E_q(J_1,J_2)}
 \left(F_q(J_1,J_2,-\gamma_2)
 - F_q(J_1,J_2,\gamma_2)\right);
 \end{eqnarray}
 \begin{eqnarray}\nonumber
 {}_q\langle J_1,\gamma_1; J_2,\gamma_2\vert \hat P_{1,q} \vert J_1,\gamma_1; J_2,\gamma_2\rangle_q &=&
 \frac{i \lambda_2\sqrt{K_1}}{2(\lambda_1 +\lambda_2)}\frac{\sqrt{J_1}}{E_q(J_1,J_2)}\left(F_q(J_1,J_2,-\gamma_1)
 - F_q(J_1,J_2,\gamma_1)\right) \\\nonumber
 &-& \frac{i \lambda_1\sqrt{K_2}}{2(\lambda_1 +\lambda_2)}\frac{\sqrt{J_2}}{E_q(J_1,J_2)}
 \left(F_q(J_1,J_2,-\gamma_2)
 - F_q(J_1,J_2,\gamma_2)\right);
 \end{eqnarray}
 \begin{eqnarray}\nonumber
 {}_q\langle J_1,\gamma_1; J_2,\gamma_2\vert \hat P_{2,q} \vert J_1,\gamma_1; J_2,\gamma_2\rangle_q &=&
 \frac{ \lambda_2\sqrt{K_1}}{2(\lambda_1 +\lambda_2)}\frac{\sqrt{J_1}}{E_q(J_1,J_2)}\left(F_q(J_1,J_2,-\gamma_1)
 - F_q(J_1,J_2,\gamma_1)\right) \\
 &+ & \frac{ \lambda_1\sqrt{K_2}}{2(\lambda_1 +\lambda_2)}\frac{\sqrt{J_2}}{E_q(J_1,J_2)}
 \left(F_q(J_1,J_2,-\gamma_2)
 - F_q(J_1,J_2,\gamma_2)\right).
\end{eqnarray}
and we have 
\begin{eqnarray}
 {}_q\langle J_1,\gamma_1; J_2,\gamma_2\vert \hat X_{1,q}^2 \vert J_1,\gamma_1; J_2,\gamma_2\rangle_q &=&
 \frac{\hbar^2 K_1}{4(\lambda_1 +\lambda_2)^2}
 \left(  1 + (1+q^2)J_1\right.\\\nonumber
 &+& \left.\frac{J_1}{E_q(J_1,J_2)}\left(F_q(J_1,J_2, -(1+q^2)\gamma_1) + 
 F_q(J_1,J_2, (1+q^2)\gamma_1)\right)\right)\\\nonumber
 &+&\frac{\hbar^2 K_2}{4(\lambda_1 +\lambda_2)^2}\left(1 + (1+ q^2)J_2\right.\\\nonumber
 &+& \left. \frac{J_2}{E_q(J_1,J_2)}\left(F_q(J_1,J_2, -(1+q^2)\gamma_2 ) 
 + F_q(J_1,J_2, (1+q^2)\gamma_2)\right)\right)\\\nonumber
 &-& \frac{\hbar^2\sqrt{K_1 K_2}}{2(\lambda_1 +\lambda_2)^2}
 \frac{(J_1J_2)^{\frac{1}{2}}}{E_q(J_1,J_2)}\left( F_q(J_1,J_2,-\gamma_1,-\gamma_2)\right.\\\nonumber
 &+&\left. F_q(J_1,J_2,\gamma_1,\gamma_2) + F_q(J_1,J_2,\gamma_1,-\gamma_2)
 + F_q(J_1,J_2,-\gamma_1,\gamma_2)\right).
 \end{eqnarray}
\begin{eqnarray}
 {}_q\langle J_1,\gamma_1; J_2,\gamma_2\vert \hat X_{2,q}^2 \vert J_1,\gamma_1; J_2,\gamma_2\rangle_q &=&
 -\frac{\hbar^2 K_1}{4(\lambda_1 +\lambda_2)^2}
 \left(  -1 - (1+q^2)J_1\right.\\\nonumber
 &+& \left.\frac{J_1}{E_q(J_1,J_2)}\left(F_q(J_1,J_2, -(1+q^2)\gamma_1) + 
 F_q(J_1,J_2, (1+q^2)\gamma_1)\right)\right)\\\nonumber
 &-&\frac{\hbar^2 K_2}{4(\lambda_1 +\lambda_2)^2}\left(-1 - (1+ q^2)J_2\right.\\\nonumber
 &+& \left. \frac{J_2}{E_q(J_1,J_2)}\left(F_q(J_1,J_2, -(1+q^2)\gamma_2 ) 
 + F_q(J_1,J_2, (1+q^2)\gamma_2)\right)\right)\\\nonumber
 &-& \frac{\hbar^2\sqrt{K_1 K_2}}{2(\lambda_1 +\lambda_2)^2}
 \frac{(J_1J_2)^{\frac{1}{2}}}{E_q(J_1,J_2)}\left( F_q(J_1,J_2,-\gamma_1,-\gamma_2)\right.\\\nonumber
 &+&\left. F_q(J_1,J_2,\gamma_1,\gamma_2) - F_q(J_1,J_2,\gamma_1,-\gamma_2)
 - F_q(J_1,J_2,-\gamma_1,\gamma_2)\right).
 \end{eqnarray}
 \begin{eqnarray}
 {}_q\langle J_1,\gamma_1; J_2,\gamma_2\vert \hat P_{1,q}^2 \vert J_1,\gamma_1; J_2,\gamma_2\rangle_q &=&
 -\frac{\lambda_2^2 K_1}{4(\lambda_1 +\lambda_2)^2}
 \left(  -1 - (1+q^2)J_1\right.\\\nonumber
 &+& \left.\frac{J_1}{E_q(J_1,J_2)}\left(F_q(J_1,J_2, -(1+q^2)\gamma_1) + 
 F_q(J_1,J_2, (1+q^2)\gamma_1)\right)\right)\\\nonumber
 &-&\frac{\lambda_1^2 K_2}{4(\lambda_1 +\lambda_2)^2}\left(-1 - (1+ q^2)J_2\right.\\\nonumber
 &+& \left. \frac{J_2}{E_q(J_1,J_2)}\left(F_q(J_1,J_2, -(1+q^2)\gamma_2 ) 
 + F_q(J_1,J_2, (1+q^2)\gamma_2)\right)\right)\\\nonumber
 &+& \frac{\lambda_1\lambda_2\sqrt{K_1 K_2}}{2(\lambda_1 +\lambda_2)^2}
 \frac{(J_1J_2)^{\frac{1}{2}}}{E_q(J_1,J_2)}\left( F_q(J_1,J_2,-\gamma_1,-\gamma_2)\right.\\\nonumber
 &+&\left. F_q(J_1,J_2,\gamma_1,\gamma_2) - F_q(J_1,J_2,\gamma_1,-\gamma_2)
 - F_q(J_1,J_2,-\gamma_1,\gamma_2)\right).
 \end{eqnarray}
 \begin{eqnarray}
 {}_q\langle J_1,\gamma_1; J_2,\gamma_2\vert \hat P_{2,q}^2 \vert J_1,\gamma_1; J_2,\gamma_2\rangle_q &=&
 \frac{\lambda_2^2 K_1}{4(\lambda_1 +\lambda_2)^2}
 \left(  1 + (1+q^2)J_1\right.\\\nonumber
 &+& \left.\frac{J_1}{E_q(J_1,J_2)}\left(F_q(J_1,J_2, -(1+q^2)\gamma_1) + 
 F_q(J_1,J_2, (1+q^2)\gamma_1)\right)\right)\\\nonumber
 &+&\frac{\lambda_1^2 K_2}{4(\lambda_1 +\lambda_2)^2}\left(1 + (1+ q^2)J_2\right.\\\nonumber
 &+& \left. \frac{J_2}{E_q(J_1,J_2)}\left(F_q(J_1,J_2, -(1+q^2)\gamma_2 ) 
 + F_q(J_1,J_2, (1+q^2)\gamma_2)\right)\right)\\\nonumber
 &+& \frac{\lambda_1\lambda_2\sqrt{K_1 K_2}}{2(\lambda_1 +\lambda_2)^2}
 \frac{(J_1J_2)^{\frac{1}{2}}}{E_q(J_1,J_2)}\left( F_q(J_1,J_2,-\gamma_1,-\gamma_2)\right.\\\nonumber
 &+&\left. F_q(J_1,J_2,\gamma_1,\gamma_2) + F_q(J_1,J_2,\gamma_1,-\gamma_2)
 + F_q(J_1,J_2,-\gamma_1,\gamma_2)\right).
 \end{eqnarray}
 Let's introduce the following functions in order to simplyfier and shorten the expressions
 \begin{equation}
  G_c(\gamma_1) = \frac{\sqrt{J_1}}{E_q(J_1,J_2)}\left( F_q(J_1,J_2,\gamma_1) + F_q(J_1,J_2,-\gamma_1)\right) 
  = \frac{2\sqrt{J_1}}{E_q(J_1,J_2)}\sum_{n_1,n_2 =0 }^\infty \frac{J_1^{n_1}J_2^{n_2}}{[n_1]_q![n_2]_q!}\cos(\gamma_1q^{2[n_1]_q});
 \end{equation}
 \begin{equation}
  G_c(\gamma_2) = \frac{\sqrt{J_2}}{E_q(J_1,J_2)}\left( F_q(J_1,J_2,\gamma_2) + F_q(J_1,J_2,-\gamma_2) \right) 
  = \frac{2\sqrt{J_2}}{E_q(J_1,J_2)}\sum_{n_1,n_2 =0 }^\infty \frac{J_1^{n_1}J_2^{n_2}}{[n_1]_q![n_2]_q!}\cos(\gamma_2q^{2[n_2]_q});
 \end{equation}
 \begin{equation}
  G_s(\gamma_1) = \frac{\sqrt{J_1}}{E_q(J_1,J_2)}\left( F_q(J_1,J_2,\gamma_1)- F_q(J_1,J_2,-\gamma_1) \right) 
  = \frac{2i\sqrt{J_1}}{E_q(J_1,J_2)}\sum_{n_1,n_2 =0 }^\infty \frac{J_1^{n_1}J_2^{n_2}}{[n_1]_q![n_2]_q!}\sin(\gamma_1q^{2[n_1]_q});
 \end{equation}
 \begin{equation}
  G_s(\gamma_2) = \frac{\sqrt{J_2}}{E_q(J_1,J_2)}\left(F_q(J_1,J_2,\gamma_2)-F_q(J_1,J_2,-\gamma_2) \right) 
  = \frac{2i\sqrt{J_1}}{E_q(J_1,J_2)}\sum_{n_1,n_2 =0 }^\infty \frac{J_1^{n_1}J_2^{n_2}}{[n_1]_q![n_2]_q!}\sin(\gamma_2q^{2[n_2]_q}),
 \end{equation}
and
\begin{equation}
 G_q(\gamma_1) = \sqrt{J_1}G_c((1+ q^2)\gamma_1);\quad 
 G_q(\gamma_2) = \sqrt{J_2}G_c((1+ q^2)\gamma_2).
\end{equation}
Moreover 
\begin{equation}
 G^+_c(\gamma_1,\gamma_2) = \frac{2\sqrt{J_1J_2}}{E_q(J_1,J_2)}\sum_{n_1,n_2 =0}^\infty
\frac{J_1^{n_1}J_2^{n_2}}{[n_1]_q![n_2]_q!}\cos(\gamma_1 q^{2[n_1]_q}+\gamma_2q^{2[n_2]_q})\,;
\end{equation}
\begin{equation}
G^+_s(\gamma_1,\gamma_2) = \frac{2 i \sqrt{J_1J_2}}{E_q(J_1,J_2)}\sum_{n_1,n_2 =0}^\infty
\frac{J_1^{n_1}J_2^{n_2}}{[n_1]_q![n_2]_q!}\sin(\gamma_1 q^{2[n_1]_q}+\gamma_2q^{2[n_2]_q})\,;
\end{equation}
\begin{equation}
G^-_c(\gamma_1,\gamma_2) = \frac{2\sqrt{J_1J_2}}{E_q(J_1,J_2)}\sum_{n_1,n_2 =0}^\infty
\frac{J_1^{n_1}J_2^{n_2}}{[n_1]_q![n_2]_q!}\cos(\gamma_1 q^{2[n_1]_q}-\gamma_2q^{2[n_2]_q})\,;
\end{equation}
\begin{equation}
G^-_s(\gamma_1,\gamma_2) = \frac{2i\sqrt{J_1J_2}}{E_q(J_1,J_2)}\sum_{n_1,n_2 =0}^\infty
\frac{J_1^{n_1}J_2^{n_2}}{[n_1]_q![n_2]_q!}\sin(\gamma_1 q^{2[n_1]_q}-\gamma_2q^{2[n_2]_q}).
 \end{equation}
With respect to equation (\ref{uncert}), we have 
\begin{eqnarray}\nonumber
 \Delta \hat X_{1,q}^2 &=&  \frac{\hbar^2K_1}{4(\lambda_1+\lambda_2)^2}
 \left(1 + (1+q^2)J_1 + G_q(\gamma_1) - G_c^2(\gamma_1)\right)\\\nonumber
 & +&  \frac{\hbar^2K_2}{4(\lambda_1+\lambda_2)^2}
 \left(1 + (1+q^2)J_2 + G_q(\gamma_2) - G_c^2(\gamma_2)\right)\\
  &+&  \frac{\hbar^2\sqrt{K_1K_2}}{2(\lambda_1+\lambda_2)^2}
 \left( -G^+_c(\gamma_1,\gamma_2) - G_c^-(\gamma_1,\gamma_2) +
 G_c(\gamma_1)G_c(\gamma_2)\right);
 \end{eqnarray}
 \begin{eqnarray}\nonumber
 \Delta \hat X_{2,q}^2 &=& \frac{\hbar^2K_1}{4(\lambda_1+\lambda_2)^2}
 \left(1 + (1+q^2)J_1 - G_q(\gamma_1) + G_s^2(\gamma_1)\right)\\\nonumber
 & +&  \frac{\hbar^2K_2}{4(\lambda_1+\lambda_2)^2}
 \left(1 + (1+q^2)J_2 - G_q(\gamma_2) + G_s^2(\gamma_2)\right)\\
  &+&  \frac{\hbar^2\sqrt{K_1K_2}}{2(\lambda_1+\lambda_2)^2}
 \left( -G^+_c(\gamma_1,\gamma_2) + G_c^-(\gamma_1,\gamma_2) +
 G_s(\gamma_1)G_s(\gamma_2)\right);
 \end{eqnarray}
 \begin{eqnarray}\nonumber
 \Delta \hat P_{1,q}^2 &=&  \frac{\lambda_2^2K_1}{4(\lambda_1+\lambda_2)^2}
 \left(1 + (1+q^2)J_1 - G_q(\gamma_1) + G_s^2(\gamma_1)\right)\\\nonumber
 & +&  \frac{\lambda_1^2K_2}{4(\lambda_1+\lambda_2)^2}
 \left(1 + (1+q^2)J_2 - G_q(\gamma_2) + G_s^2(\gamma_2)\right)\\
  &+&  \frac{\lambda_1\lambda_2\sqrt{K_1K_2}}{2(\lambda_1+\lambda_2)^2}
 \left( G^+_c(\gamma_1,\gamma_2) - G_c^-(\gamma_1,\gamma_2) 
 -G_s(\gamma_1)G_s(\gamma_2)\right);
\end{eqnarray}
\begin{eqnarray}\nonumber
 \Delta \hat P_{2,q}^2 &=& 
  \frac{\lambda_2^2K_1}{4(\lambda_1+\lambda_2)^2}
 \left(1 + (1+q^2)J_1 + G_q(\gamma_1) - G_c^2(\gamma_1)\right)\\\nonumber
 & +&  \frac{\lambda_1^2K_2}{4(\lambda_1+\lambda_2)^2}
 \left(1 + (1+q^2)J_2 + G_q(\gamma_2) - G_c^2(\gamma_2)\right)\\
  &+&  \frac{\lambda_1\lambda_2\sqrt{K_1K_2}}{2(\lambda_1+\lambda_2)^2}
 \left( G^+_c(\gamma_1,\gamma_2) + G_c^-(\gamma_1,\gamma_2) 
 - G_c(\gamma_1)G_c(\gamma_2)\right).
 \end{eqnarray}
 In the general case, means $\gamma_i \neq 0, i=1,2$, in order 
 for the equation (\ref{gur}) to be satisfied for all simultaneous 
 measurement, the following inequalities should hold.
 \begin{eqnarray}\nonumber
 \left( K_1\left[2J_1 + G_q(\gamma_1)\right]\right. 
 &+& K_2\left[2 J_2 + G_q(\gamma_2)\right]
 -2\sqrt{K_1K_2}\left[G^+_c(\gamma_1,\gamma_2) +G^-_c(\gamma_1,\gamma_2)\right]\\\label{p1}
  &-&\left. \left[\sqrt{K_1}G_c(\gamma_1)-\sqrt{K_2}G_c(\gamma_2)\right]^2 \right) \ge 0;
\end{eqnarray}
\begin{eqnarray}\nonumber
 \left( K_1\left[2J_1 - G_q(\gamma_1)\right] \right.
 &+& K_2\left[2 J_2 - G_q(\gamma_2)\right]
 -2\sqrt{K_1K_2}\left[G^+_c(\gamma_1,\gamma_2) - G^-_c(\gamma_1,\gamma_2)\right]\\\label{p2}
 &+& \left.\left[\sqrt{K_1}G_s(\gamma_1)+ \sqrt{K_2}G_s(\gamma_2)\right]^2 \right) \ge 0
\end{eqnarray}
\begin{eqnarray}\nonumber
 \left( \lambda_2^2 K_1\left[2J_1 - G_q(\gamma_1)\right]\right. 
 &+& \lambda_1^2 K_2\left[2 J_2 - G_q(\gamma_2)\right]
 +2\lambda_1\lambda_2 \sqrt{K_1K_2}\left[G^+_c(\gamma_1,\gamma_2) - G^-_c(\gamma_1,\gamma_2)\right]\\\label{p3}
 &+&\left.\left[\lambda_2\sqrt{K_1}G_s(\gamma_1)- \lambda_1\sqrt{K_2}G_s(\gamma_2)\right]^2\right) \ge 0
\end{eqnarray}
\begin{eqnarray}\nonumber
 \left(\lambda_2^2 K_1\left[2J_1 + G_q(\gamma_1)\right]\right. 
 &+& \lambda_1^2 K_2\left[2 J_2 + G_q(\gamma_2)\right]
 +2\lambda_1\lambda_2 \sqrt{K_1K_2}\left[G^+_c(\gamma_1,\gamma_2) + G^-_c(\gamma_1,\gamma_2)\right]\\\label{p4}
 &-&\left. \left[\lambda_2\sqrt{K_1}G_s(\gamma_1) + \lambda_1\sqrt{K_2}G_s(\gamma_2)\right]^2 \right) \ge 0
\end{eqnarray}

{\bf Acknowledgments:} L. Gouba is supported by the Abdus Salam International Centre for Theoretical Physics (ICTP).

\end{document}